\documentclass[prb,aps,epsf,twocolumn,superscriptaddress,showpacs,10pt]{revtex4-2}
\usepackage{graphicx,amsfonts,times,bm,amsmath,verbatim,color,array}
\usepackage{lineno}
\usepackage{xcolor}
\usepackage{algorithmic}
\usepackage{bbold}
\usepackage{braket}
\usepackage[colorlinks,urlcolor=blue,citecolor=blue,linkcolor=blue]{hyperref}
\usepackage{siunitx}

\begin{document}

\title{Electric field induced Berry curvature dipole and non-linear anomalous Hall effects in higher-wave symmetric unconventional magnets}
\author{Srimayi Korrapati}
\affiliation{Department of Physics and Astronomy, Clemson University, Clemson, SC 29634, USA}
\author{Snehasish Nandy}
\affiliation{Department of Physics, National Institute of Technology Silchar, Assam, 788010, India}
\author{Sumanta Tewari}
\affiliation{Department of Physics and Astronomy, Clemson University, Clemson, SC 29634, USA}

\begin{abstract}
We investigate the second-order anomalous Hall response in two-dimensional higher-wave symmetric magnets, including the recently discovered class of collinear magnets known as altermagnets, when subjected to a symmetry-breaking external electric field. In these systems, the first- and second-order anomalous Hall responses mediated by the first- and second-order multipoles of the Berry curvature over the occupied states vanish by symmetry. However, a symmetry-breaking dc electric field can induce a nonzero Berry curvature dipole by coupling to a non-vanishing quantum metric, also known as the Berry connection polarizability. An applied ac electric field can then generate a finite nonlinear transverse Hall effect characterized by a second harmonic response. In addition, the dc field itself can generate a finite third order transverse Hall response. We discuss this remarkable effect in a class of higher-order symmetric unconventional magnets (of $p$, $d$, $f$, $g$, $i$ symmetry), including the subclass of altermagnets. We demonstrate that the electric-field-induced anomalous Hall effect in the higher-wave-symmetric magnets can serve not only as a probe of the underlying quantum metric of the occupied states but also as a means to distinguish the even ($d$-,$g$-wave) and odd ($p$-wave) order parameter symmetries defined on the square lattice.
\end{abstract}

\maketitle
\section{\label{seclevel1}Introduction} A recently discovered class of collinear magnets, collectively referred to as higher-wave-symmetric unconventional magnets, exhibits the intriguing coexistence of momentum-dependent spin splitting and vanishing net magnetization~\cite{Kusunose_2019, Zunger_2020, Libor_2021, Jungwirth_2022, Tomas_2022}. In these systems, spin polarization arises from non-relativistic crystalline symmetries, enabling spin splitting even in the absence of spin-orbit coupling (SOC). These unconventional magnets, with $p, d, f, g, i$-wave symmetric spin splittings,  are characterized by unique band structures in which the number of nodes is $1, 2, 3, 4,$ and $6$, respectively~\cite{Zunger_2020, Libor_2021, Jungwirth_2022, Ezawa_2025}. In particular, the $d$-, $g$-, and $i$-wave magnets belong to the subclass known as altermagnets. The altermagnets break time-reversal symmetry (TRS) but preserve inversion symmetry. In contrast, $p$-wave and $f$-wave magnets break inversion symmetry while respecting TRS. The zero net magnetization with, combined with the spin-split band structure of higher-wave symmetric magnets, sets them apart from both conventional ferromagnets ($s$-wave) and antiferromagnets, while simultaneously posing challenges for experimental detection using standard magnetic probes. Nevertheless, their symmetry-protected spin-split band structures give rise to a rich landscape of unconventional transport and optical phenomena, opening new avenues for spintronics, quantum information, and correlated electron systems~\cite{Yao_2024, Lorenzo2025, Kriegner_2023, Libor_2022, Nandy_2025, Hariki_2024, Antonenko_2024, Farajollahpour_2025, Jungwirth2025, Song_2025, Korrapati_2025, Chakraborti_2025, Ezawa_2025}.


The Hall effect, characterized by the emergence of a transverse voltage in response to a longitudinal current, has played a pivotal role in unveiling topological phases of matter and driving advances in electronic and spintronic technologies. Among the family of Hall effects, the linear intrinsic anomalous Hall effect (AHE), observable even in the absence of external magnetic fields, serves as a direct probe of the Berry curvature, the imaginary component of the quantum geometric tensor (QGT), which constitutes a cornerstone of modern topological band theory~\cite{Nagaosa_2010, Xiao_2010}. However, the linear AHE necessitates TRS breaking due to Onsager reciprocity relations~\cite{Nagaosa_2010}. In two-dimensional (2D) higher-wave-symmetric magnets, the linear intrinsic AHE vanishes in altermagnets despite the TRS breaking as a direct consequence of their underlying crystalline symmetries and in $p$-wave and $f$-wave magnets due to the presence of TRS. For instance, in $d$-wave altermagnets, the combined $\hat{C}_{4z}\hat{\mathcal{T}}$ ($\hat{C}_{4z}$ is the four-fold rotational symmetry around the $z$-axis) symmetry enforces this cancellation. Under this symmetry, the Berry curvature transforms as $\Omega(k_x,k_y) \rightarrow -\Omega(-k_y,k_x)$, leading to the exact cancellation of $\Omega(k_x,k_y)$ and $\Omega(-k_y,k_x)$ upon Brillouin zone integration, thereby nullifying the net Berry curvature monopole~\cite{Korrapati_2025}.

Beyond linear response, the Berry curvature dipole (BCD), defined as the first moment of the Berry curvature over occupied states, governs the second-order AHE, giving rise to a second-harmonic Hall voltage under an ac longitudinal current~\cite{Sodemann_2015}. This nonlinear effect underlies potential applications in high-frequency rectifiers, energy harvesters, and infrared detectors~\cite{Du_2021, ZDu_2021}. However, the BCD is severely constrained by crystal symmetry: in two dimensions, it survives only in systems with a single mirror-line~\cite{Sodemann_2015, Nandy_2019, ZDu_2021}. Interestingly, the BCD also vanishes in 2D higher-wave-symmetric  magnets due to symmetries. For example, it has been shown that in $d$-wave altermagnets, despite the breaking of both $\hat{C}_{4z}$ and $\hat{\mathcal{T}}$ individually, the derivative $\partial{k_a}\Omega(k_x,k_y)$ transforms as $-\partial_{k_a}\Omega(-k_x,-k_y)$, making it odd in momentum. The resulting Brillouin zone integral therefore vanishes, suppressing second-order responses and yielding a dominant third-order anomalous Hall effect dictated by the underlying quantum geometry~\cite{Fang_2024}.

Recently, an alternative route to realize a finite BCD has been proposed through the concept of Berry connection polarizability~\cite{Gao_2014}, which captures the field-induced positional shift of Bloch electrons~\cite{Ye_2023, Liu_2021, Ankita_2025}. A dc electric field can generate a field-induced BCD via the nontrivial quantum metric of occupied bands, enabling a symmetry-allowed second-order anomalous Hall effect, even in systems where the intrinsic BCD vanishes. This mechanism not only circumvents stringent symmetry constraints for a non-zero BCD, but also provides a tunable platform for controlling nonlinear Hall responses. 

In this paper, we show that a dc electric field can induce a non-zero Berry curvature dipole through its coupling with the quantum metric in higher-wave-symmetric magnets, including the subclass of altermagnets. The induced BCD enables a second-order anomalous Hall response when an ac field is applied, manifesting as a second-harmonic response. In addition, the dc field itself can generate a finite third order anomalous Hall response. We consider $p$, $d$, $f$, $g$, and $i$-wave symmetric magnets and calculate the behavior of the induced BCD and the resultant non-linear conductivities. We then analyze the experimentally relevant angular dependence of the second- and third-order anomalous Hall conductivities, considering variations with respect to the in-plane angles $\phi$ and $\theta$, defined as the in-plane angles of the dc and ac electric fields relative to a mirror line, respectively. Note that $f$ and $i$-wave systems are defined on a triangular lattice, while the rest of the unconventional magnets are defined on the square lattice. We find that, among unconventional magnets defined on the square lattice, the induced BCD and the nonlinear Hall response are qualitatively similar for $d$ and $g$ magnets, while they show distinct behavior for $p$-wave magnets. For example, as shown in Fig. \ref{fig:d_wave}(e), the induced BCD is always perpendicular to the dc electric field in the $d$-wave magnets, independent of $\phi$. Conversely, as shown in Fig. \ref{fig:p_wave}(e), in the $p$-wave magnets, the induced BCD and the applied dc field are not always perpendicular to each other. 
The third order dc anomalous Hall response is finite and shows a $\sin2\phi$ dependence in $p$-wave magnets, but is forbidden by symmetry in $d$-wave magnets.  Additionally, $f$ and $i$-wave magnets defined on the triangular lattice show similar behavior to $d$-wave magnets. Thus, the electric-field-induced second-order anomalous Hall effect in higher-wave symmetric magnets can serve not only as a probe of the underlying quantum geometry of the Bloch states but also as a means to distinguish between even ($d$, $g$-wave) and odd ($p$-wave) parity order parameter symmetries defined on the square lattice.

The remainder of the paper is organized as follows: In Sec.~\ref{sec: Theory}, we describe the mechanism by which an electric field induces a Berry curvature dipole in systems possessing a non-trivial quantum metric of the occupied bands. Sec.~\ref{sec: Hamiltonian} introduces the model Hamiltonians and provides the expressions for the Berry connection polarizability (BCP) in two-band systems. Sec.~\ref{sec: Results} presents our results for various higher-wave-symmetric unconventional magnets, followed by a brief summary and conclusions in Sec.~\ref{sec: Discussions and conclusions}.

\section{\label{sec: Theory} Electric Field Induced Berry Curvature Dipole}
In this section, we demonstrate that a dc electric field can induce a finite Berry curvature dipole even in systems where it is otherwise prohibited by crystalline symmetries. The emergence of this finite BCD originates from the field-induced correction to the Berry curvature. A dc electric field $\bm{E}^{\rm{dc}}$ is introduced as a perturbative correction to the Hamiltonian as 
\begin{equation}\label{eq:perturbation}
    {H}^{\prime}_E = -e\bm{E}^{\rm{dc}}\cdot(\bm{r}-\bm{r}_{c}),
\end{equation}
where $\bm{r}_c$ denotes the center of the Bloch wave packet. The field causes the wave packet to acquire a positional shift relative to its center, captured by the first-order correction to the cell-periodic part of the Bloch eigenstate~\cite{Gao_2014, Liu_2021, Nag_2023}:
\begin{eqnarray}
|u^{(1)}_{m k}\rangle=\sum_{n \neq m}\frac{|u^{(0)}_{n k}\rangle\langle u^{(0)}_{n k}|{H}^{\prime}_E|u^{(0)}_{m k}\rangle}{\epsilon^{(0)}_{mn, k}}=\sum_{n \neq m}\frac{e\bm{E}^{\rm{dc}}\cdot \bm{\mathcal{A}}^{(0)}_{nm}|u^{(0)}_{nk}\rangle}{\epsilon^{(0)}_{mn, k}}, \nonumber 
\label{en_corr}
\end{eqnarray}
where the position operator acts as $\bm{r}=i\partial_{\bm{k}}$, $\epsilon^{(0)}_{mn, \bm{k}}=\epsilon_{m \bm{k}}^{(0)}-\epsilon_{n \bm{k}}^{(0)}$, and $m,  n$ are band indices. Here, ${\epsilon}_{m k}^{(0)}$ and $\bm{\mathcal{A}}_{m n}^{(0)}$ are the unperturbed band energy and interband Berry connection, respectively, where $\bm{\mathcal{A}}_{m n}^{(0)}$ can be expressed as $\bm{\mathcal{A}}_{m n}^{(0)}=\langle u^{(0)}_{m k}|i\bm{\nabla_{\bm{k}}}|u^{(0)}_{n k}\rangle$. The first-order correction to the Berry connection, $A_{m,a}^{(1)}$, measuring a shift in its center of mass position of the wave packet, gives the positional shift for the band $m$ as
\begin{eqnarray}
A_{m,a}^{(1)}&
=&2e{\rm Re}\sum_{n}^{n \neq m}\frac{A^{(0)}_{m n,a} A^{(0)}_{nm,b}}{\epsilon_{m k}^{(0)}-\epsilon_{n k}^{(0)}}E_b=eG_{m,ab}(E^{\rm{dc}})_b,
\label{BC_corr}
\end{eqnarray}
where $G_{m,ab}$, termed the Berry connection polarizability, is a gauge-invariant tensor, and
$a$, $b$ represent Cartesian coordinates. Analogous to electric polarizability in electrodynamics, which quantifies the tendency of matter to develop an electric dipole moment under an external field, the BCP characterizes the shift of Bloch electrons in momentum space, thereby inducing a Berry curvature dipole.

We emphasize the relationship between BCP ($G_{m,ab}$) and the quantum metric $\mathcal{Q}_{m,ab}$, given their significant contributions to the nonlinear Hall effect. The quantum metric tensor ($\mathcal{Q}_{m,ab}= \text{Re}\sum_{m'\neq m}  \bm{\mathcal{A}}_{m m'}^{(0)} \bm{\mathcal{A}}_{m' m}^{(0)} $) is closely connected to the interband Berry connection ($\bm{A}_{m m'}^{(0)}$) associated with unperturbed states. Remarkably, the BCP is revealed to be a band-renormalized quantity of the quantum metric, expressed succinctly as 
\begin{equation}\label{eq: BNQM}
    G_{m,ab}=2\text{Re}\sum_{m'\neq m}  \frac{\bm{A}_{m m'}^{(0)} \bm{A}_{m'm}^{(0)}}{\epsilon^{(0)}_{mm^\prime, k}},
\end{equation}
highlighting its dependence on interband energy separations~\cite{Torma_2023, Barman_2025}.

The BCP generates a first-order (in dc electric field ${E}_{dc}$) correction to the Berry curvature \cite{Gao_2014}:
 \begin{equation}\label{eq:ber_cor}
    \begin{split}
        \bm{\Omega}^{(1)}_{\bm{k}} =e&\nabla_{\bm{k}} \times (\overleftrightarrow{G}\cdot {\bm{E}^{\rm{dc}}}).\\
 \end{split}
\end{equation}
In 2D, the Berry curvature is a pseudoscalar with only the out-of-plane component being non-zero and independent; we write $\bm{\Omega}_{\bm{k}}\equiv\Omega_{\bm{k},z}$ and omit the subscript $z$ in the remainder of this work for simplicity. The corresponding 2D BCD pseudovector components $\mathcal{D}_{bz}\equiv\mathcal{D}_{b}$ can then be obtained as \cite{Sodemann_2015}
\begin{equation}
\mathcal{D}^{(1)}_{a}=\int_{\bm{k}}f_0(\partial_a\Omega^{(1)}_{\bm{k}}),
\end{equation}
where $f_0$ is the equilibrium Fermi-Dirac distribution function and 
\begin{equation}\label{eq:BCmoment}
        \begin{split}     \partial_a\Omega^{(1)}_{\bm{k}} 
        =&eE^{\rm{dc}}\cos\phi(\partial^2_{xa} G_{xy} - \partial^2_{ay} G_{xx}) \\&+eE^{\rm{dc}}\sin\phi(\partial^2_{xa} G_{yy} - \partial^2_{ay} G_{xy}), 
    \end{split}
\end{equation}
with $\phi$ being the angle between the $x$-direction (or more generally a mirror line) and the applied dc field. The highest symmetry permitting a nonzero BCD in a 2D system is a single mirror line, which constrains the BCD to lie perpendicular to the mirror line \cite{Sodemann_2015}. In systems having higher symmetries that forbid a finite BCD, an in-plane dc electric field lowers the symmetry to at most one of the existing mirror lines, supporting a nonzero BCD in the perturbed system \cite{Ye_2023}. The symmetry is lowered because  $H^{\prime}_{E}$ (Eq.~(\ref{eq:perturbation})) preserves only the mirror line of the unperturbed system that is parallel to  $\bm{E}^{\rm{dc}}$, while all others are broken.

In a 2D system with RSOC, the presence of a mirror line symmetry in the unperturbed system guaranties that the field-induced BCD is perpendicular to the applied field, both when $\bm{E}^{\rm{dc}}$ is parallel and when it is perpendicular to the mirror line. The symmetry argument is as follows. Suppose there exists a mirror-line symmetry along the $x$-direction ($\hat{\mathcal{M}_{y}}:k_x\rightarrow-k_{x}$). The components of the BCP (see Eqs.~ (\ref{eq:QGT}) and (\ref{eq:GAB})) then transform as  $G_{aa}(k_{x},k_y)\rightarrow G_{aa}(-k_{x},k_y)$ and $G_{ab}(k_{x},k_y)\rightarrow-G_{ab}(-k_{x},k_y)$, where $a\neq b$. As a result, the mixed second order derivatives ($\partial^{2}_{xy}$) of $G_{aa}$  and pure second order derivatives ($\partial^{2}_{xx},\partial^2_{yy}$) of $G_{ab}$ (with $a\neq b$) are both odd under $k_x$. Since the integration over the Brillouin zone spans a symmetric interval, the integrals of these odd terms vanish in Eq.~(\ref{eq:BCmoment}).
The orientation of the BCD pseudovector in the transport plane is then dictated by 
\begin{equation}\label{eq:BCDcomponents}
\begin{split}    
\mathcal{D}^{(1)}_{x}(\phi)=&eE^{\rm{dc}}\sin\phi\tilde{\mathcal{D}}_{x},\\
\mathcal{D}_{y}^{(1)}(\phi)=&
eE^{\rm{dc}}\cos\phi\tilde{\mathcal{D}}_{y},
\end{split}
\end{equation}
where $\phi$ is the angle of the dc field measured from the mirror line, and
\begin{equation}\label{eq:BCDamp}
\begin{split}    
\tilde{\mathcal{D}}_{x}=&\int_{\bm{k}}f_{0}(\partial^2_{xx} G_{yy} - \partial^2_{xy} G_{xy}),\\
\tilde{\mathcal{D}}_{y}=&\int_{\bm{k}}f_{0}(\partial^2_{xy} G_{xy} - \partial^2_{yy} G_{xx})
\end{split}
\end{equation}
are the amplitudes of the BCD components (in units of e$E^{\rm{dc}}$).
Eq.~(\ref{eq:BCDcomponents}) indicates that the $\mathcal{D}^{(1)}_{x}$ is determined solely by the $y$-component of the dc field, while $\mathcal{D}^{(1)}_{y}$ is determined solely by the $x$-component of the same. Therefore, the induced BCD is orthogonal to the dc electric field whenever $\bm{E}^{\rm{dc}}$ is oriented parallel to or perpendicular to a mirror line that is a symmetry of the unperturbed system. However, as we show in Fig.~\ref{fig:p_wave}(b), when the dc electric field makes an arbitrary angle (i.e., $\phi \neq 0,\pi/2$) relative to the mirror line, the induced BCD is not necessarily perpendicular to $\bm{E}^{\rm{dc}}$. As we discuss in the following sections, this marks an important distinction between the $p$-wave (odd parity) and $d,g$-wave (even parity) symmetric unconventional magnets defined on the square lattice. 

The field-induced BCD can be detected through the second harmonic ($2\omega$) response driven by a probing ac field, $E^{\omega}_{a}(t)=\text{Re}\{\mathcal{E}_ae^{i\omega t} \}$, satisfying $E^{\omega}<<E^{\rm{dc}}$. Including the dc field induced correction to the Berry curvature, the equation of motion for the wavepacket center in the presence of a driving ac field can be written as 
\begin{equation}\label{eq:velocity}
\dot{{\bm{r}}}=\frac{1}{\hbar}\nabla_{\bm{k}}{\epsilon}_{\bm{k}}+\frac{e}{\hbar}\bm{E}{^{\omega}}\times(\bm{\Omega}^{(0)}+\bm{\Omega}^{(1)}).
\end{equation}
The charge current is given by
\begin{equation}
    j_a = -e\int_{\bm{k}} f(\epsilon_{\bm{k}},t) \dot{r}_a,
\end{equation}
where the non-equilibrium distribution function $f(\epsilon_{\bm{k}},t)$ is obtained from the semiclassical Boltzmann transport equation under the relaxation time approximation as 
\begin{equation}
\dot{\bm{k}}.\nabla_{\bm{k}} + \frac{\partial f}{\partial t}=\frac{f_0(\epsilon_{\bm{k}})-f(\epsilon_{\bm{k}},t)}{\tau}.\\
\end{equation}
Here, we assume the driving electric field oscillates harmonically in time but is uniform in space, and $\tau$ is the relaxation time. We ignore the momentum dependence of the $\tau$ for simplicity. Since we are interested in computing the second-order Hall conductivity, expanding $f=\text{Re}\{f_0+f_1+f_2\}$ in powers of $E^{\omega}$ such that $f_\nu=\sum f_{\nu}^{\nu\omega}e^{i\nu\omega t}$, the first-order term is \cite{Sodemann_2015}
\begin{equation}
\begin{split}
    f_1=f_1^{\omega}e^{i\omega t},~~~~~ f_1^\omega=&\frac{e}{\hbar}\frac{\partial_af_0}{\tilde{\omega}}\mathcal{E}_{a},  
\end{split} 
\end{equation}
where $\tilde{\omega}=i\omega+1/\tau$.
The second-order Hall current, with a rectified component $j_{a}^0$ and a second-harmonic component $j_{a}^{2\omega}$, is obtained as $j_a^{(2)} = \text{Re}\{j^0_a+j^{2\omega}_ae^{2i\omega t}\}$
where
\begin{equation}\label{eq: secondOrderCurrent}
\begin{split}
j^0_a=\chi_{abc}\mathcal{E}_b\mathcal{E}_c^*,~j^{2\omega}_a=\chi_{abc}\mathcal{E}_b\mathcal{E}_c,
\end{split}
\end{equation}
where the components of the 2D second-order Hall conductivity tensor are obtained to be directly proportional to the BCD components as 
\begin{equation}\label{eq:chi}
\chi_{abc}=\varepsilon_{ac}\frac{e^3}{\hbar^2}\frac{\tau}{ 2(1+i\omega\tau)} \mathcal{D}^{(1)}_{b}. 
\end{equation}
Here, the contribution from $\mathcal{D}_{b}^{(0)}$ has been dropped since the intrinsic BCD vanishes due to crystal symmetry in the  higher-wave symmetric systems under investigation. Notice that the nonlinear Hall conductivity tensor $\chi_{abc}$ is symmetric under the last two indices. To isolate the dissipationless components, $\chi_{abc}$ must be antisymmetric with respect to the first
index, either with the second or third, which  are equivalent by construction. Therefore, in 2D, the independent components of the second-order Hall conductivity tensor can be identified as $\chi_{xyy} = -\chi_{yyx}$ and $\chi_{yxx} = -\chi_{xxy}$.

Having identified the components of the field induced dipole and second-order Hall conductivity, we can formulate the experimentally accessible transverse second-order Hall conductivity $\chi^{(2\omega)}=j^{2\omega}/\mathcal{E}^2=(\chi_{yxx}\cos\theta-\chi_{xxy}\sin\theta)$ \cite{Ye_2023,Liu_2021}. Here, $\chi^{(2\omega)}$ is defined such that it directly relates the amplitude $\mathcal{E}$ of the probing ac field (directed at an angle $\theta$ measured from a mirror line) to the second harmonic transverse response (along $\theta+\pi/2$) given by  
\begin{equation}\label{eq:secondOrderCurrent_vectors}
j^{2\omega}=\frac{e^3\tau}{ 2(1+i\omega\tau)\hbar^2} (\hat{\bm{z}}\times\bm{E}^{\omega})(\boldsymbol{\mathcal{D}}^{(1)}(\phi)\cdot\bm{E}^{\omega}).
\end{equation} 
In this work, where the BCD is induced by a symmetry-lowering dc field, we express the transverse second-order anomalous Hall conductivity as
\begin{equation}\label{eq:secondOrderAC}
\chi^{(2\omega)}(\phi,\theta)=\frac{e^3\tau}{ 2(1+i\omega\tau)\hbar^2}(\mathcal{D}^{(1)}_{x}(\phi)\cos\theta+\mathcal{D}^{(1)}_{y}(\phi)\sin\theta).
\end{equation}
We note that with $\mathcal{D}^{(1)}\propto E^{\rm{dc}}$, $\chi^{(2\omega)}$ also scales linearly with $E^{\rm{dc}}$. In addition to the second order ac Hall current, the field-induced BCD could, in general, give rise to a non-linear dc anomalous Hall  response driven by the symmetry-lowering dc field itself:
\begin{equation}\label{eq:j3}
j^{(3)}=\frac{e^3\tau}{ \hbar^2} (\hat{\bm{z}}\times\bm{E}^{\rm{dc}})(\boldsymbol{\mathcal{D}}^{(1)}(\phi)\cdot\bm{E}^{\rm{dc}}).
\end{equation}
This is a straightforward result obtained by treating $E_{dc}$ as the driving field with frequency $\omega=0$ in Eqs.~(\ref{eq:velocity}-\ref{eq:secondOrderCurrent_vectors}). Here, we recognize that the induced BCD $\boldsymbol{\mathcal{D}}^{(1)}$ itself is linear in $E^{\rm{dc}}$, making the current a third order response \cite{Lai2021,Nag_2023},
$j^{(3)}_{a}=\chi^{(3,\rm{dc})}_{abcd}E^{\rm{dc}}_{b}E^{\rm{dc}}_{c}E^{\rm{dc}}_{d}$. 
For 2D systems with RSOC and at least one mirror line, the transverse conductivity corresponding to the third-order anomalous Hall current generated in response to a dc electric field is then: 
\begin{equation}\label{eq:thirdOrderDC}
\chi^{\rm{dc}}(\phi)=\frac{e^4}{\hbar^2}\tau(\tilde{\mathcal{D}}_{x}+\tilde{\mathcal{D}}_{y})\sin2\phi,
\end{equation}
where $\phi$ is the angle between $\bm{E}^{\rm{dc}}$ and the mirror line, and we have used Eqs.~(\ref{eq:BCDcomponents}) and (\ref{eq:BCDamp}) in Eq.~(\ref{eq:j3}). Note that while $\mathcal{D}^{(1)}_{a}$ is linearly proportional to $E^{\rm{dc}}$, $\tilde{\mathcal{D}}_{a}$ is not, making $\chi^{\rm{dc}}$ independent of $E^{\rm{dc}}$. As we discuss in the following sections, this third order response is finite only in the case of the $p$-wave magnet under consideration, which has a single mirror plane and no rotational symmetry; here, the field-induced BCD is not strictly perpendicular to the in-plane $\bm{E}^{\rm{dc}}$ for all angles $\phi$. We do not include a discussion of the third order responses due to the inherent Berry curvature quadrupole which scale as $\tau^2$ \cite{PhysRevB.107.115142}, and limit the discussion to the nonlinear responses mediated by the field induced BCD which scale as $\tau$, as seen in Eqs.~(\ref{eq:secondOrderAC}) and (\ref{eq:thirdOrderDC}). 

\section{\label{sec: Hamiltonian}Model Hamiltonian of Unconventional Magnets}
In this section, we examine the Hamiltonians of higher-wave symmetric magnets with $p,d,f,g$ or $i$ wave symmetry. The single orbital generic lattice Hamiltonian of higher-wave-symmetric magnets in the presence of Rashba spin-orbit coupling (RSOC) can be written as
\begin{eqnarray}\label{eq:hamiltonian}
  &\mathcal{H}(\bm{k}) = \mathcal{H}_{kin}(\bm{k}) + \mathcal{H}_{RSOC} (\bm{k})+ q_{\mathbf{k}}\sigma_z\;,
\end{eqnarray}
where the first term of $\mathcal{H}_{kin}$ represents the kinetic energy term, the second term $\mathcal{H}_{RSOC}$ denotes the RSOC term, and $q_{\mathbf{k}}$ is the momentum dependent spin-splitting form factor of the unconventional magnetic order parameter. The lattice constant
is taken to be unity without any loss of generality. It is important to note that the Hamiltonian for $p, d, g$-wave magnets is written on the square lattice, whereas for $f$-wave and $i$-wave magnets, the Hamiltonians are described on a triangular lattice. In the case of $p, d, g$-wave magnets, the kinetic energy and RSOC terms take the following forms:
\begin{eqnarray}
    &\mathcal{H}_{kin} = -2t(\cos k_x+\cos k_y),\nonumber \\
    &\mathcal{H}_{RSOC}=\lambda (\sin k_{y} \sigma_{x} - \sin k_{x} \sigma_{y}),
\end{eqnarray}
where $t$ is the nearest-neighbor hopping parameter, and $\lambda$ is the strength of the Rashba spin-orbit coupling. At a finite $t$, the Dirac nodes at $\Gamma=(0,0)$ and $\rm{M}=
(\pi,\pi)$ become separated in energy in these square-lattice  Hamiltonians. Meanwhile, the kinetic and RSOC terms for $f$ and $i$-wave magnets defined on the triangular lattice take the following forms \cite{Ezawa_2025}:
\begin{equation}
\begin{split}
     \mathcal{H}_{kin} = 2t&\bigg(3-\cos k_x - \cos \frac{k_{x}+\sqrt{3}k_y}{2}- \cos \frac{k_{x}-\sqrt{3}k_y}{2}\biggr) \nonumber
\end{split}
\end{equation}
\begin{equation}
    \begin{split}
        \mathcal{H}_{RSOC}=&\frac{2}{3}\lambda\bigg(\sin k_x + \sin \frac{k_x}{2}\cos \frac{\sqrt{3}k_y}{2}\biggr)\sigma_y\\&-\frac{2\sqrt{3}}{3}\lambda\cos{\frac{k_x}{2}}\sin\frac{\sqrt{3}k_y}2 \sigma_x.
    \end{split}
\end{equation}

The imaginary (Berry curvature $\Omega^{\pm}_{z}$) and real (quantum  metric) part of the quantum geometric tensor for a two-band system described as $tk^2+\bm{d_{\bm{k}} \cdot \sigma}$ can be calculated as \cite{PhysRevB.104.085114}
\begin{equation}\label{eq:QGT}
    \begin{split}
        \Omega^{\pm}_{z} &=\mp\frac{1}{2}\hat{\bm{d_k}}\cdot(\partial_x\hat{\bm{d_k}}\times\partial_y\hat{\bm{d_k}}) \\
        g^{\pm}_{ab} &= \frac{1}{4}\partial_a\hat{\bm{d_k}}\cdot \partial_b\hat{\bm{d_k}}, \\
    \end{split}
\end{equation}
where $\pm$ represent the conduction and valence bands, respectively, with the corresponding energies being
\begin{equation}
    \epsilon^{\pm}_{\bm{k}}=tk^2\pm d_{\bm{k}}.
\end{equation}
It is important to note that the Berry curvature is fully antisymmetric and the quantum metric is solely symmetric. The Berry connection polarizability tensor, also known as the band-normalized quantum metric, is given by 
\begin{equation}\label{eq:GAB}
    G_{ab}^{\pm}=\mp\frac{g_{ab}}{2d_{\bm k}}.
\end{equation}

 \section{\label{sec: Results} Results} 
We investigate the Berry curvature dipole that arises from the Berry connection polarizability when unconventional magnets are subjected to a dc electric field. 
While we explicitly discuss $p$-wave and $d$-wave models in this work, we find that the results for $g$-, $f$- and $i$-wave order parameters are similar to the $d$-wave case, while the $p$-wave case is distinct.
\begin{figure*}[!t]
\centering
\includegraphics[width=0.98\textwidth]{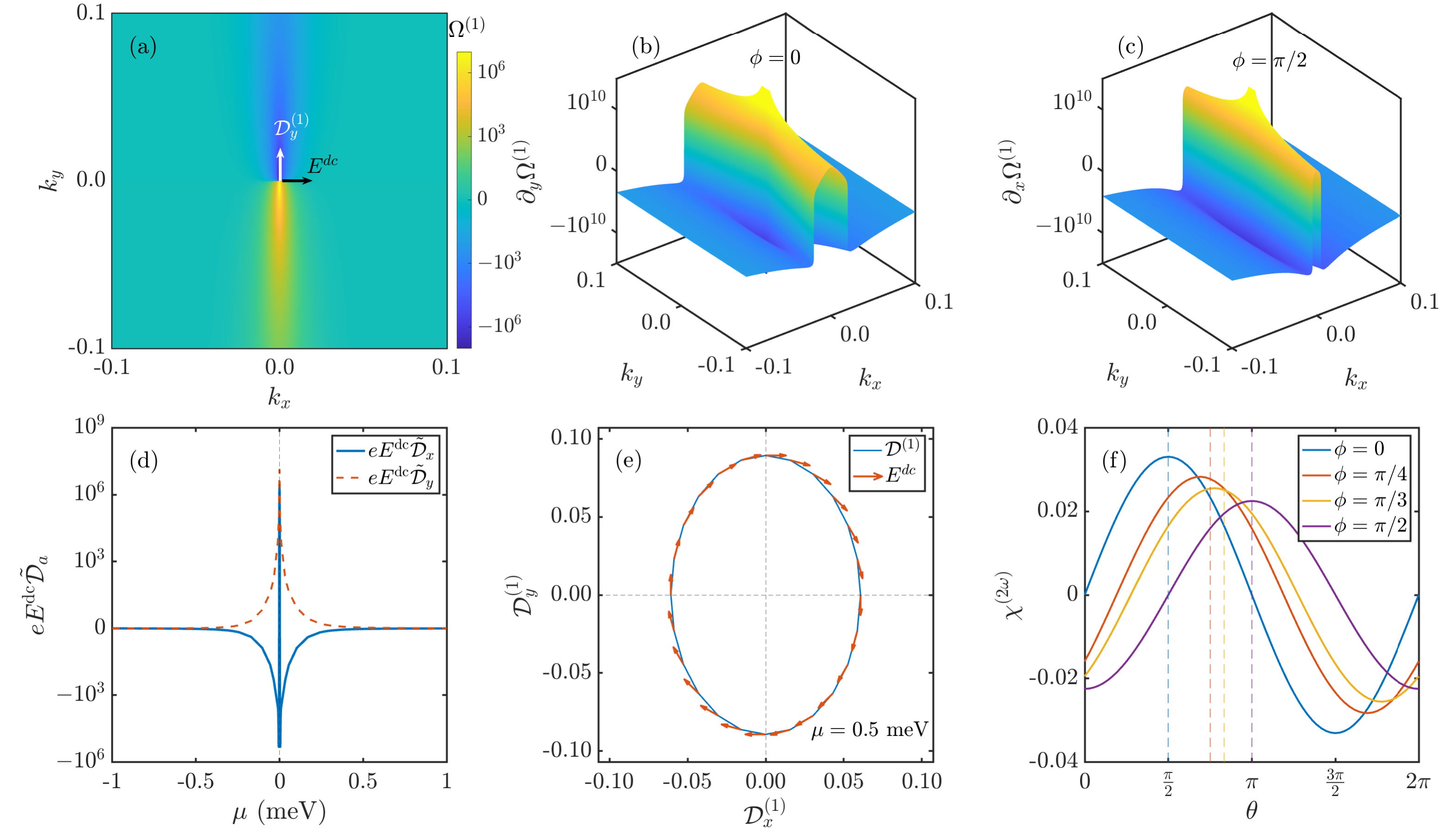}
\caption{BCD induced by a dc electric field for a $p$-wave magnet (Eq.~(\ref{eq:hamiltonian}) with form factor Eq.~(\ref{eq:p-wave})) with model parameters $t=1$ eV, $\lambda=0.1\,t$,  $\Delta_{p}=0.5\,t$, $E^{\rm{dc}}=3$~kV/m, scattering time $\tau=10^{-12}~\si{\second}$, and lattice constant $a= 4~\text{\AA}$. (a) The first-order correction to Berry curvature $\Omega^{(1)}$ in the presence of $\bm{E}^{\rm{dc}}$ along the $x$-direction, i.e. $\phi=0$, where the induced BCD is perpendicular to the dc field. The dipolar nature of the distribution is evident with mirrored positive and negative regions. BCD distributions (b) $\partial_{y}\Omega^{(1)}$ with $\bm{E}^{\rm{dc}}$ along $\phi = 0$, and (c)  $\partial_{x}\Omega^{(1)}$ with $\bm{E}^{\rm{dc}}$ along $\phi = \pi/2$, illustrating that the two cases are not equivalent. This accounts for the anisotropy in the magnitudes of $\tilde{\mathcal{D}}_{x}$ and $\tilde{\mathcal{D}}_{y}$. (d) Unequal amplitudes of $\boldsymbol{\mathcal{D}}^{(1)}$ (see Eq.~(\ref{eq:BCDamp})) along $x$ and $y$ directions (in the log scale) as a function of the chemical potential varied near the band-touching point at $\mu=0$. (e) Polar plot of the field-induced BCD $\boldsymbol{\mathcal{D}}^{(1)}$ showing directional anisotropy. The direction of the symmetry-reducing electric field ${\bm{E}^{\rm{dc}}}$ is indicated by orange arrows, and the induced $\boldsymbol{\mathcal{D}}^{(1)}$ is understood to be the vector starting at the origin and ending at the base of the corresponding arrow. For the $p$-wave system under consideration, $\boldsymbol{\mathcal{D}}^{(1)}$ is only perpendicular to $\bm{E}^{\rm{dc}}$ when the applied dc field is along the $x$- or $y$-directions. As ${\bm{E}^{\rm{dc}}}$ rotates clockwise, $\boldsymbol{\mathcal{D}}^{(1)}$ rotates in the same sense for the chemical potential chosen. (f) The second-order Hall conductivity $\chi^{2\omega}$ (Eq.~(\ref{eq:secondOrderAC})) versus the angle $\theta$ (w.r.t. the $x$-direction) of the probing field $\bm{E}^{\omega}$ for various  $\bm{E}^{\rm{dc}}$ orientations $\phi$, with the chemical potential set to $\mu=0.5$~meV. The vertical dashed lines mark the expected angular positions of the maximal values of $\chi^{(2\omega)}$ for each $\phi$ (denoted by the corresponding color) in an isotropic system. The directional anisotropy of $\boldsymbol{\mathcal{D}}^{(1)}$ is manifest in $\chi^{(2\omega)}$. The units for $\Omega^{(1)}$ are $\text{\AA}^2$, those of $\mathcal{D}^{(1)}$ are $\si{\nano\meter}$, $\chi^{2\omega}$ has units of $\si{\volt^{-1}\siemens\nano\meter}$.}
\label{fig:p_wave}
\end{figure*}
We begin with the $p_x$-wave model, where the form factor in Eq.~(\ref{eq:hamiltonian}) is given by
\begin{equation}\label{eq:p-wave}
    q_{\bm{k}}^{\rm{p_x-wave}}=\Delta_{p}\sin k_x,
\end{equation} 
with $\Delta_{p}$ being the strength of the unconventional $p$-wave order. In the context of symmetries, the $p$-wave system preserves $\hat{\mathcal{T}}$ (TRS) in addition to the mirror-line symmetry $\hat{\mathcal{M}}_{y}:k_x \rightarrow -k_x$. The spin-texture in this system, specified by $\bm{\hat{d}_{k}}$ with $\bm{d_k}~=~(\lambda\sin k_y,-\lambda\sin k_x,\Delta_{p}\sin k_x)$, lies entirely in a single plane, and the resultant Berry curvature (Eq.~(\ref{eq:QGT})) is identically zero over the Brillouin zone (except at the Dirac points). However, the application of $\bm{E}^{\rm{dc}}$ allows the system to acquire a field-induced correction to the Berry curvature through the BCP. We analytically obtain the components of the BCP tensor using the low-energy Hamiltonian of the $p$-wave system near the $\Gamma$-point:
\begin{equation}\label{eq:p_lin}
    \mathcal{H}^{\rm p-wave} = tk^2+\lambda k_y\sigma_x-\lambda k_x \sigma_y + \Delta_{p}k_x\sigma_z,
\end{equation}
where the $\bm{k}$-independent term has been omitted, as it contributes only a constant energy shift.
The components of the BCP tensor for the valence band are obtained as
 \begin{equation}\label{eq:p_BCP}
    \begin{split}          G_{xx}^{-}=&\frac{\lambda^2}{8d_{\bm{k}}^5}(\Delta_{p}^2 + \lambda^2)k_y^2,\quad     G_{yy}^{-}=\frac{\lambda^2}{8d_{\bm{k}}^5}(\Delta_{p}^2 + \lambda^2)k_x^2,\\
     G_{xy}^{-}=&G_{yx}^{-}=-\frac{\lambda^2}{8d_{\bm{k}}^5}(\Delta_{p}^2 + \lambda^2)k_xk_y,
    \end{split}
\end{equation}
where $d_{\bm{k}}=\sqrt{\lambda^2(k_x^2+k_y^2)+\Delta_{p}^2k_x^2}$ is the energy splitting between the bands, from the $\bm{d_k}$ vector of Eq.~(\ref{eq:p_lin}). The field-induced Berry curvature $\Omega^{(1)}$ (Eq.~(\ref{eq:ber_cor})) when $\bm{E}^{\rm{dc}}$ is oriented along the $x$-direction ($\phi=0$) is depicted in Fig.~\ref{fig:p_wave}(a).  The distribution of $\Omega^{(1)}$ features mirrored positive and negative regions, revealing its dipolar nature and suggesting a nonzero field induced BCD $\boldsymbol{\mathcal{D}}^{(1)}$. The distribution of $\partial_{y}\Omega^{(1)}$ with $\bm{E}^{\rm{dc}}$ along $\phi = 0$ determines the amplitude $\tilde{\mathcal{D}}_{y}$ of the $y$-component of the BCD (see Eqs.~(\ref{eq:BCmoment} -\ref{eq:BCDamp})) and is visualized in Fig.~\ref{fig:p_wave}(b). In Fig.~\ref{fig:p_wave}(c), we show the distinct distribution of $\partial_{x}\Omega^{(1)}$ with $\bm{E}^{\rm{dc}}$ along $\phi = \pi/2$, which determines the amplitude $\tilde{\mathcal{D}}_{x}$.
A comparison of the two is indicative of an anisotropy in the amplitudes of $\boldsymbol{\mathcal{D}}^{(1)}$ in the $x$- and $y$-directions, consistent with Fig.~\ref{fig:p_wave}(d), which shows the variation of $\tilde{\mathcal{D}}_{x}$ and $\tilde{\mathcal{D}}_{y}$ as a function of chemical potential ($\mu$). While the two profiles differ across the entire range of $\mu$, both components exhibit a peak (albeit with different magnitudes) at the band touching point $\mu=0$, as expected from the dependence of the BCD components on the interband energy separations (see Eq.~\ref{eq: BNQM}). To gain further insight into this anisotropy, we examine Eq.~(\ref{eq:BCDamp}), which reveals that both $\tilde{\mathcal{D}}_{x}$ and $\tilde{\mathcal{D}}_{y}$ have a contribution from $\partial_{xy}G_{xy}$. Any anisotropy between their values, therefore, originates from an inequality between $\partial_{xx}G_{yy}$ and $\partial_{yy}G_{xx}$. Although the BCP terms in Eq.~(\ref{eq:p_BCP}) appear deceptively symmetric under the exchange $k_x\leftrightarrow k_y$, the anisotropy is concealed in the denominator via $d_{\bm{k}}$ corresponding to a $p$-wave form factor. Consequently, for a 2D system with Rashba coupling and $p$-wave form factor, $\int_{k}f_{0}\partial_{yy}G_{xx}\neq\int_{k}f_{0}\partial_{xx}G_{yy}$. Using this inequality along with Eqs.~(\ref{eq:BCDcomponents}) and (\ref{eq:BCDamp}) confirms that $\tilde{\mathcal{D}}_{x}\neq-\tilde{\mathcal{D}}_{y}$, and it immediately follows that for the $p$-wave form factor in Eq.~(\ref{eq:p-wave}),
\begin{equation}\label{eq:notperp}
\boldsymbol{\mathcal{D}}^{(1)}\cdot\bm{E}^{\rm{dc}}=e(E^{\rm{dc}})^2(\tilde{\mathcal{D}}_{x}+\tilde{\mathcal{D}}_{y})\sin 2\phi\neq0
 \end{equation}
where we find that the induced BCD is perpendicular to the field $\bm{E}^{\rm{dc}}$ only when $\phi=0,\pi/2$; this is in agreement with the discussion involving mirror symmetries above Eq.~(\ref{eq:BCDcomponents}) in Sec.~\ref{sec: Theory}. The  anisotropy is further highlighted in the polar plot of $\boldsymbol{\mathcal{D}}^{(1)}$ in Fig.~\ref{fig:p_wave}(e), which shows that both the magnitude and orientation of $\boldsymbol{\mathcal{D}}^{(1)}$ vary upon the rotation of the applied field $\bm{E}^{\rm{dc}}$ in the plane. From an experimental standpoint, Eq.~(\ref{eq:notperp}) means that the driving ac field and the BCD-inducing dc electric field need not be perpendicular to each other to obtain the maximal current response for $\phi\neq0,\pi/2$. From Eqs. (\ref{eq: secondOrderCurrent}) and (\ref{eq:chi}), we see that a component $\mathcal{E}_{a}$ of a driving field produces a perpendicular second-order anomalous Hall current $j_{\perp a}$ through a parallel component of BCD: $\mathcal{D}^{(1)}_a$. Accordingly, the current response is maximized when the ac field is parallel to the induced dipole, i.e.,  $\bm{E}^{\omega}\parallel\boldsymbol{\mathcal{D}}^{(1)}$.  
From Eq.~(\ref{eq:notperp}) and  Fig.~\ref{fig:p_wave}(e), we gather that for $\phi=0,\pi/2$, the maximal second-order ac Hall response is generated by the following configuration: $\boldsymbol{\mathcal{D}}^{(1)}\perp\bm{E}^{\rm{dc}}$,  $\bm{E}^{\omega}\parallel\boldsymbol{\mathcal{D}}^{(1)}$, and $ \bm{E}^{\omega}\perp\bm{E}^{\rm{dc}}$. However, for $\phi\neq 0,\pi/2$, $\boldsymbol{\mathcal{D}}^{(1)}$ is not purely perpendicular to the dc electric field. Then, the driving ac field that generates the maximal current response, being parallel to the BCD, is also not purely perpendicular to $\bm{E}^{\rm{dc}}$. Fig.~\ref{fig:p_wave}(f) shows that as the angle $\theta$ of the driving $\bm{E}^{\omega}$ is varied at a fixed $\phi$ (where $\phi\neq 0,\pi/2$), the maximal current response occurs at an angular position shifted away from its expected value (indicated by the vertical dashed lines) of $\phi+\pi/2$ . We also note that the anisotropy between the BCD components along $x$- and $y$- directions manifests as different maximal amplitudes of $\chi^{2\omega}$ in Fig.~\ref{fig:p_wave}(f) for different orientations $\phi$ of the BCD inducing dc field.

\begin{figure}[!t]
\centering
\includegraphics[width=0.42\textwidth]{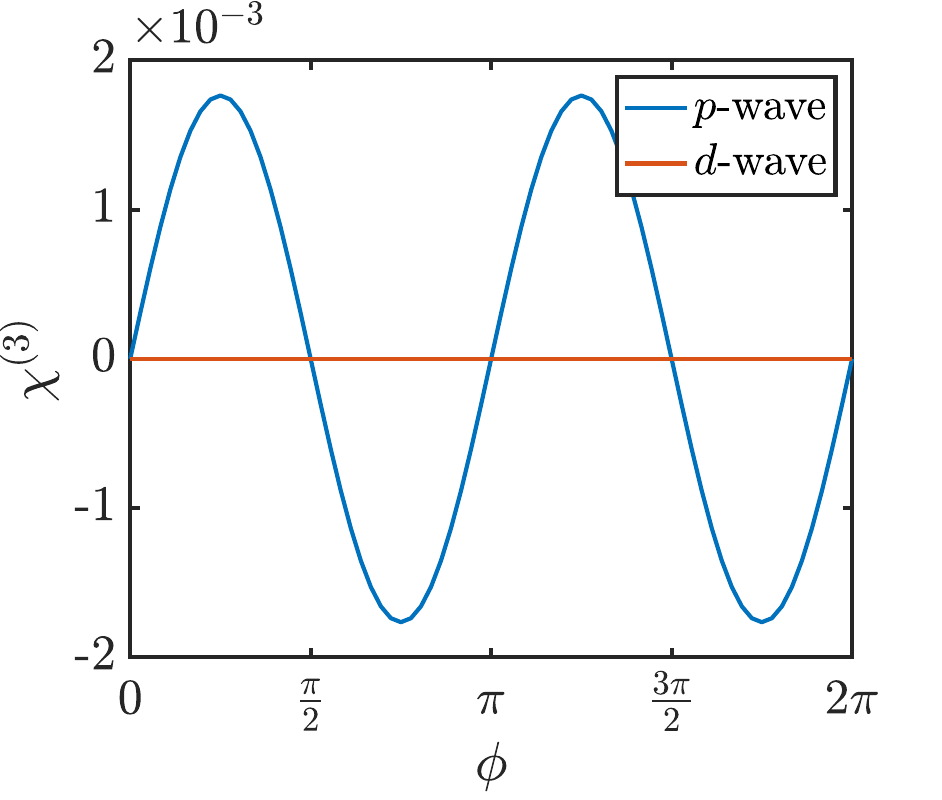}
\caption{Third-order dc anomalous Hall conductivity $\chi^{(\rm{dc})}$ (see Eq.~(\ref{eq:thirdOrderDC})) originating from field-induced BCD for a Hamiltonian (Eq.~(\ref{eq:hamiltonian}), showing a $\sin{2\phi}$ dependence for a $p$-wave form factor (Eq.~(\ref{eq:p-wave})), and vanishing for a $d$-wave form factor (Eq.~(\ref{eq:d-wave})). Units for $\chi^{(3)}$ are $\si{\volt^{-2}\siemens\micro\meter^{2}}$. The model parameters are the same as the ones mentioned in the caption of Fig.~\ref{fig:p_wave}.}
\label{fig:thirdOrder}
\end{figure}

Another remarkable experimental consequence of Eq.~(\ref{eq:notperp}) is that $\bm{E}^{\rm{dc}}$ itself could drive a third-order dc current $j^{(3)}=\chi^{(3)}(E^{\rm{dc}})^3$ (see Eq.~(\ref{eq:j3})) by coupling to a parallel component of the induced BCD. The predicted $\sin(2\phi)$ dependence of $\chi^{(3)}$ in Eq.~(\ref{eq:thirdOrderDC}) is understood as the $\sin(2\phi)$ dependence in Eq.~(\ref{eq:notperp}) of the component of BCD parallel to the dc field. The predicted angular dependence of  $\chi^{(3)}$ is shown in Fig.~\ref{fig:thirdOrder}, with the response vanishing whenever $\boldsymbol{\mathcal{D}}^{(1)}\perp\bm{E}^{\rm{dc}}$, i.e., when the dc field makes an angle $\phi=0,\pi/2$ with the mirror plane along $x$-direction. while $\bm{E}^{\omega}$ drives both a dc $j^{(0)}$ and a second harmonic ac $j^{(2\omega)}$ response in the second-order (see Eq.~(\ref{eq: secondOrderCurrent})). $j^{(2\omega)}$ is experimentally distinguishable using frequency lock-in techniques. With $E^{\omega}<<E^{\rm{dc}}$, the third-order dc response dominates since $j^{(0)}/j^{(3)}\propto (E^{\omega}/E^{\rm{dc}})^2<<1$. The third order dc Hall response is independent of the ac field. 
 \begin{figure*}[!htb]
\centering
\includegraphics[width=0.98\textwidth]{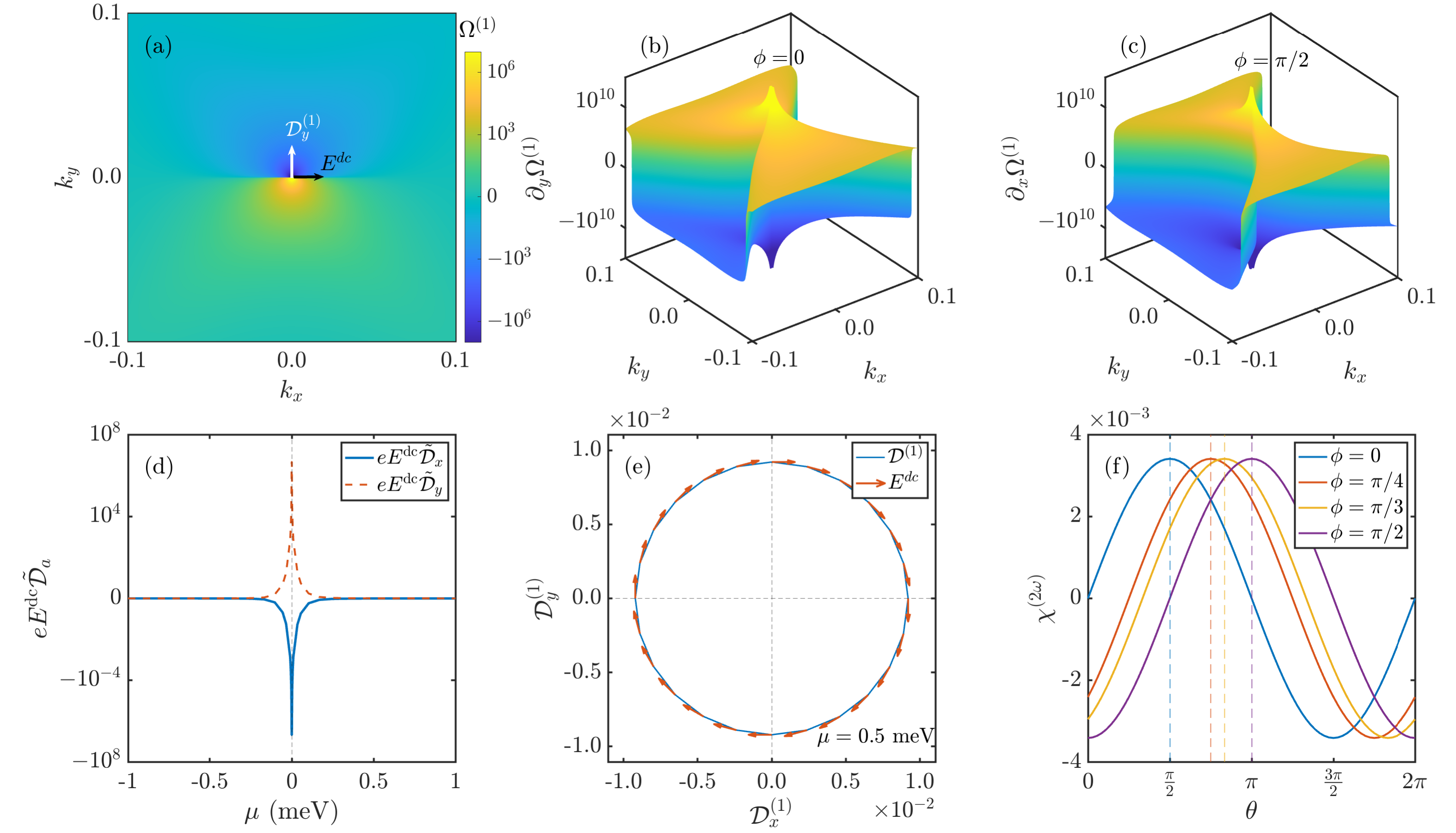}
\caption{BCD induced by dc electric field for a $d$-wave altermagnet for which the form factor is Eq.~(\ref{eq:d-wave}) with model parameters $t=1$ eV, $\lambda=0.1\,t$,  $\Delta_{p}=0.5\,t$, $E^{\rm{dc}}=3$~kV/m, scattering time $\tau=10^{-12}~\si{\second}$ and lattice constant $a= 4~\text{\AA}$. (a) The first-order correction to Berry curvature $\Omega^{(1)}$ in the presence of $\bm{E}^{\rm{dc}}$ along the $x$-direction, i.e., $\phi=0$, where the induced BCD is perpendicular to the dc field. The dipolar nature of the distribution is evident with mirrored positive and negative regions. BCD distributions (b) $\partial_{y}\Omega^{(1)}$ with $\bm{E}^{\rm{dc}}$ along $\phi = 0$, and (c)  $\partial_{x}\Omega^{(1)}$ with $\bm{E}^{\rm{dc}}$ along $\phi = \pi/2$, illustrating that the two cases are equivalent. This explains the isotropy in the amplitudes of of $\mathcal{D}^{(1)}_{x}$ and $\mathcal{D}^{(1)}_{y}$. (d) Equal amplitudes of $\boldsymbol{\mathcal{D}}^{(1)}$ (see Eq.~\ref{eq:BCDamp}) along $x$ and $y$ directions (in the log scale) as a function of the chemical potential varied near the band-touching point at $\mu=0$. (e) Polar plot of the field-induced BCD $\boldsymbol{\mathcal{D}}^{(1)}$ showing directional isotropy. The direction of the symmetry-reducing electric field ${\bm{E}^{\rm{dc}}}$ is indicated by orange arrows, and the induced $\boldsymbol{\mathcal{D}}^{(1)}$ is understood to be the vector starting at the origin and ending at the base of the corresponding arrow. For the $d$-wave system under consideration, $\boldsymbol{\mathcal{D}}^{(1)}$ remains perpendicular to $\bm{E}^{\rm{dc}}$ as the field is rotated in the plane. As ${\bm{E}^{\rm{dc}}}$ rotates clockwise, $\boldsymbol{\mathcal{D}}^{(1)}$ rotates in the same sense for any chemical potential. (f) The second-order Hall conductivity $\chi^{2\omega}$ (Eq.~(\ref{eq:secondOrderAC})) versus the angle $\theta$ (w.r.t. the $x$-direction) of the probing field $\bm{E}^{\omega}$ for various  $\bm{E}^{\rm{dc}}$ orientations $\phi$, with the chemical potential set to $\mu=0.5$ meV. The vertical dashed lines mark the expected angular positions of the maximum value of $\chi^{(2\omega)}$ for each $\phi$ (denoted by the corresponding color) in an isotropic system. The directional isotropy of the BCD is manifest in $\chi^{(2\omega)}$. The units for $\Omega^{(1)}$ are $\text{\AA}^2$, those of $\mathcal{D}^{(1)}$ are $\si{\nano\meter}$, $\chi^{2\omega}$ has units of $\si{\volt^{-1}\siemens\nano\meter}$.}
\label{fig:d_wave}
\end{figure*}

The discussion now proceeds to $d$-wave altermagnets. In the general case, a $d$-wave altermagnet can host both $d_{x^2 - y^2}$ and $d_{xy}$ order parameters with respective strengths $\Delta_{d}$ and $\Delta_{d}^{\prime}$, where The $d_{x^2 - y^2}$ term typically originates from nearest-neighbor hopping, while the $d_{xy}$ term arises from second-nearest-neighbor hopping on the square lattice~\cite{Farajollahpour_2025}. The two symmetries are related to each other by a $\pi/4$ rotation. In this work, we set $\Delta_{d}^{\prime}=0$ and focus exclusively on the $d_{x^2 - y^2}$-wave altermagnet. The corresponding form factor for the Hamiltonian Eq.~(\ref{eq:hamiltonian}) can be written as
\begin{equation}\label{eq:d-wave}
    q_{\bm{k}}^{\rm d-wave}=\frac{\Delta_{d}}{2}(\cos k_x-\cos k_y).
\end{equation} 
In the presence of Rashba SOC, the Hamiltonian for the $d$-wave altermagnet breaks $\hat{\mathcal{P}}$, $\hat{\mathcal{T}}$, and $\hat{\mathcal{C}}_{4z}$ symmetries. The relevant preserved symmetries are $\hat{\mathcal{C}}_{4z}\hat{\mathcal{T}}$, $\hat{\mathcal{M}}_{x=y}$ and $\hat{\mathcal{M}}_{x=-y}$. The $\mathcal{\hat{C}}_{4z}\hat{\mathcal{T}}$ enforces a vanishing first-order response. The presence of two mirror line symmetries results in a vanishing intrinsic BCD, thereby suppressing a second-order Hall response in the absence of a symmetry lowering field. When the system is subjected to $\bm{E}^{\rm{dc}}$, a field-induced Berry curvature is generated via the BCP and exhibits a dipolar nature.  We analytically evaluate the components of the BCP tensor using the low-energy Hamiltonian expanded near the $\Gamma$-point, given by 
\begin{equation}\label{eq:d_lin}
    \mathcal{H}^{\rm d-wave} = tk^2+\lambda k_y\sigma_x-\lambda k_x \sigma_y + \Delta_{d}(k_y^2-k_x^2)\sigma_z,
\end{equation}
where the $\bm{k}$-independent term has been dropped. The components of the BCP for the valence band are obtained as
\begin{equation}
    \begin{split}               G_{xx}^{-}=&\frac{\lambda^2}{8d_{\bm k}^5}(\Delta_{d}^2k_x^4 + 6\Delta_{d}^2k_x^2k_y^2 + \Delta_{d}^2k_y^4 + k_y^2\lambda^2)\\     G_{yy}^{-}=&\frac{\lambda^2}{8d_{\bm k}^5}(\Delta_{d}^2k_x^4 + 6\Delta_{d}^2k_x^2k_y^2 + \Delta_{d}^2k_y^4 + k_x^2\lambda^2)\\
     G_{xy}^{-}=&\frac{-k_xk_y\lambda^2}{8d_{\bm k}^5}(4\Delta_{d}^2k_x^2 + 4\Delta_{d}^2k_y^2 + \lambda^2)
    \end{split}
\end{equation}
where $d_{\bm k}=\sqrt{\lambda^2(k_x^2+k_y^2)+\Delta_{d}^2(k_y^2-k_x^2)^2}$ is the energy splitting between the bands, from the $\bm{d_{k}}$ vector of Eq.~(\ref{eq:d_lin}). The field-induced Berry curvature $\Omega^{(1)}$ (see Eq.~($\ref{eq:ber_cor}$)) in the presence of $\bm{E}^{\rm{dc}}$ applied along $x$ ($\phi=0$) is depicted in Fig.~\ref{fig:d_wave}(a). The distribution of $\Omega^{(1)}$ features mirrored positive and negative regions, revealing its dipolar nature and suggesting a nonzero field induced BCD $\boldsymbol{\mathcal{D}}^{(1)}$. The distributions of $\partial_{y}\Omega^{(1)}$ with $\bm{E}^{\rm{dc}}$ along $\phi = 0$, and of $\partial_{x}\Omega^{(1)}$ with $\bm{E}^{\rm{dc}}$ along $\phi = \pi/2$ are visualized in Figs.~\ref{fig:d_wave}(b) and (c), respectively. These equivalent distributions, related by an in-plane $\pi/2$ rotation and a sign reversal, allude to the amplitudes of the induced dipole components being equal in magnitude and opposite in sign. This is precisely what is seen in Fig.~\ref{fig:d_wave}(d), which shows that the profiles of the absolute amplitudes of both components overlap as the chemical potential ($\mu$) is varied, with an expected divergence at the band touching point $\mu=0$ from Eq.~(\ref{eq: BNQM}). The predicted isotropy arises from the $\hat{\mathcal{C}}_{4z}\hat{\mathcal{T}}$ symmetry of the system, which enforces $\int_{k}f_{0}\partial_{yy}G_{xx}=\int_{k}f_{0}\partial_{xx}G_{yy}$. It follows from Eq.~(\ref{eq:BCDcomponents}) that the resultant amplitudes of BCD in $x$- and $y$-directions are equal and opposite in a $d$-wave system, i.e., $\tilde{\mathcal{D}_{x}}=-\tilde{\mathcal{D}_{y}}$. 
Using this equality along with Eqs.~(\ref{eq:BCDcomponents}) and (\ref{eq:BCDamp}), it immediately follows that in the case of the $d$-wave form factor ,
\begin{equation}
\boldsymbol{\mathcal{D}}^{(1)}\cdot\bm{E}^{\rm{dc}}=e(E^{\rm{dc}})^2(\mathcal{D}^{(1)}_{0x}+\mathcal{D}^{(1)}_{0y})\sin 2\phi=0,
 \end{equation}
 suggesting that the dc field always induces a BCD perpendicular to itself. The polar plot of $\boldsymbol{\mathcal{D}}^{(1)}$ in Fig.~\ref{fig:d_wave}(d) illustrates the persistent orthogonality between the symmetry-lowering dc field and the induced BCD. 
 This directional locking is in sharp contrast to the angular dependence of $\boldsymbol{\mathcal{D}}^{(1)}$ for $p$-wave systems (Fig. \ref{fig:p_wave}(b))  and is a direct result of the isotropic BCD amplitudes being equal in magnitude and opposite in sign. The first experimental consequence of the orthogonality condition $\boldsymbol{\mathcal{D}}^{(1)}\cdot\bm{E}^{\rm{dc}}=0$ is that the maximal second-harmonic response, which requires $\bm{E}^{\omega}\parallel\boldsymbol{\mathcal{D}}^{(1)}$ (see Eq.~ (\ref{eq:secondOrderCurrent_vectors})), is obtained when $\bm{E}^{\omega}\perp\bm{E}^{\rm{dc}}$ . This is in agreement with Fig.~\ref{fig:d_wave}(f) where we investigate the dependence of $\chi^{2\omega}$ on the in-plane angle $\theta$ (measured from the $x$-axis) of the driving field $\bm{E}^{\omega}$ at various orientations $\phi$ of $\bm{E}^{\rm{dc}}$. Here, the peaks, which correspond to the maximum current response, always occur at $\theta=\phi+\pi/2$ (as indicated by the vertical dashed lines of colors corresponding to the value of $\phi$), making $\bm{E}^{\omega}\perp\bm{E}^{\rm{dc}}$. In addition, with only the component of $\bm{E}^{\omega}$ that is parallel to $\boldsymbol{\mathcal{D}}^{(1)}$ contributing to a second-harmonic, second-order Hall response transverse to $\bm{E}^{\omega}$, a configuration with $\bm{E}^{\rm{dc}}\parallel \bm{E}^{\omega}$ cannot produce a response governed by the field-induced BCD for any in-plane angle of $\bm{E}^{\rm{dc}}$ with the constraint  $\boldsymbol{\mathcal{D}}^{(1)}\perp\bm{E}^{\rm{dc}}$. This too is in agreement with Fig.~\ref{fig:d_wave}(f) where we see $\chi^{2\omega}$ crossing zero at $\theta=\phi$ for all values of $\phi$. The other experimental consequence is that with $\boldsymbol{\mathcal{D}}^{(1)}\cdot\bm{E}^{\rm{dc}}=0$, $\bm{E}^{\rm{dc}}$ is unable to drive a third order current as seen from Eq.~(\ref{eq:j3}), and $\chi^{(3)}(\phi)=0$; the third-order dc Hall response is suppressed in the case of a $d$-wave order parameter and is indicated in Fig.~\ref{fig:thirdOrder}.

We briefly discuss the non-linear anomalous Hall effects in other higher-wave symmetric unconventional magnets. First, we look at the $g$-wave magnet with the form factor
\begin{equation}\label{eq:g-wave}
    q_{\bm{k}}^{\rm g-wave} = \Delta_{g}k_xk_y(k_x^2-k_y^2),
\end{equation}
and note that the full unperturbed Hamiltonian (Eq. (\ref{eq:hamiltonian})) is invariant under four mirror line symmetries: $\hat{\mathcal{M}}_{x},\hat{\mathcal{M}}_{x},\hat{\mathcal{M}}_{x=y},\hat{\mathcal{M}}_{x=-y}$. The intrinsic Berry curvature itself is nonzero, but the unperturbed BCD is vanishing due to the four mirror lines of symmetry. Thus, the second order response to a probing ac field will be solely from the dc field induced BCD. As discussed in connection with Eq.~(\ref{eq:BCDcomponents}), when $\bm{E}^{\rm{dc}}$ is directed along a mirror line of symmetry, the induced BCD is constrained to lie perpendicular to it. In addition, the Hamiltonian is invariant under $\hat{\mathcal{C}}_{4z}$ symmetry, making the amplitudes of the induced BCD components equal in magnitude and opposite in sign ($\tilde{\mathcal{D}}_{x}=\tilde{\mathcal{D}}_{y}$) , as discussed in the $d$-wave case, ensuring that the BCD remains perpendicular to the symmetry-lowering field as $\bm{E}^{\rm{dc}}$ is rotated within the plane. Consequently, the second order in ac response is isotropic and the third order in dc response is vanishing, like in the case of the $d$-wave altermagnet. We now turn our focus to the $f$ and $i$-wave magnets defined on a triangular lattice, where the low-energy expressions of the form factors become
\begin{equation}\label{eq:f-wave}
    q_{\bm{k}}^{\rm f-wave} = \Delta_{f} k_x(3k_y^2-k_x^2),
\end{equation}
\begin{equation}\label{eq:i-wave}
    q_{\bm{k}}^{\rm i-wave} = \Delta_{i}k_xk_y(3k_y^2-k_x^2)(k_y^2-3k_x^2).
\end{equation}  
where $\Delta_{f}$ and $\Delta_{i}$ are the strengths of the $f$-wave and $i$-wave magnetic orders. While $f$-wave magnets have odd-order form factors like the $p$-wave magnet, the behavior of the induced BCD is qualitatively similar to the even order parameters of the $d$-, $g$ and $i$-wave magnets. The $f$- and $i$- wave systems have a nonzero intrinsic Berry curvature, as in the case of the $d$- and $g$-wave systems. However, the intrinsic BCD vanishes in these systems due to multiple mirror lines of symmetry: $\hat{\mathcal{M}}_{x},\hat{\mathcal{M}}_{x=3y},\hat{\mathcal{M}}_{x=-3y}$ in the case of $f$-wave, and $\hat{\mathcal{M}}_{x},\hat{\mathcal{M}}_{y},\hat{\mathcal{M}}_{x=3y},\hat{\mathcal{M}}_{x=-3y},\hat{\mathcal{M}}_{3x=y},\hat{\mathcal{M}}_{3x=-y}$ in the case of $g-$wave. We have calculated the field-induced Berry curvature and corresponding BCD and find that they remain qualitatively similar to the $d$-wave case described above, since any one of the mirror symmetries constrains the BCD to be perpendicular to the field when it is applied along their mirror lines. The $C_{3}$ symmetry that exists in both of these systems ensures that the amplitudes of the BCD components along the $x$- and $y$- directions are equal (and opposite) for any in-plane orientation of the dc field, thereby constraining the induced BCD to remain orthogonal to the dc field as it is rotated in the plane. In general, a 2D system with RSOC having at least one mirror line symmetry, in addition to an $n$-fold rotational symmetry about the out-of-plane $z$-axis with $n>2$, would exhibit an isotropic field-induced BCD. The anomalous Hall responses associated with the field-induced BCD are thereby expected to be  qualitatively similar to the case of the $d$-wave altermagnet, making the $p$-wave magnetic order distinct.

The variation of the field-induced BCD with respect to an applied dc electric field provides a powerful experimental probe for distinguishing between $d$- and $p$-wave symmetric unconventional magnets. Based on the discussions  corresponding to the $p$-wave and $d$-wave order parameters, we arrive at the following conclusions.
For $d$-wave altermagnets, the driving ac electric field must possess a component perpendicular to the dc electric field in order to generate a second-order anomalous Hall response transverse to the ac field through nonzero second-harmonic generation. In contrast, for $p$-wave unconventional magnets, the second-order Hall response remains finite even when the ac and dc electric fields are parallel, provided that the dc field is applied at an angle $\phi \neq 0,\pi/2$ measured from the $x$-axis (or more generally, a mirror line). Moreover, the field-induced BCD in $p$-wave magnets gives rise to a third-order dc Hall response, which is absent in $d$-wave systems.
The anisotropy between the two components of the BCD in the $p$-wave magnet leads to a corresponding anisotropy in the second-order Hall conductivity.  As the ac field is rotated within the plane, the expected angular dependence of the second-order Hall conductivity $\chi^{2\omega}$ (Eq.~(\ref{eq:secondOrderAC})) for various orientations of the dc electric field in $p_x$-wave and $d_{x^2-y^2}$-wave magnets is shown in Figs.~\ref{fig:p_wave}(f) and \ref{fig:d_wave}(f), respectively. $\chi^{2\omega}$, which is linear in $E^{\rm{dc}}$, can be experimentally determined by applying a dc field to induce a BCD and measuring the second-harmonic response generated perpendicular to a weak probing ac field ($E^{\rm{dc}} \gg E^{\omega}$). In the absence of an ac field, as the dc field is rotated within the plane, the expected angular dependence of the third-order dc anomalous Hall conductivity $\chi^{(3)}$ (Eq.~(\ref{eq:thirdOrderDC})), which is finite in the case of $p$-wave magnet and vanishes in the case of $d$-wave magnet, is shown in Fig.~\ref{fig:thirdOrder}. $\chi^{(3)}$, which does not scale with $E^{\rm{dc}}$, can be experimentally determined by applying a dc field to induce a BCD and measuring the third order dc response generated perpendicular to $E^{\rm{dc}}$ itself. These distinct angular profiles directly reflect the anisotropy of the Berry curvature dipole, providing a practical means to distinguish between even- and odd-parity orders of unconventional altermagnetism on a square lattice.

\section{\label{sec: Discussions and conclusions}Summary and Conclusions}

In summary, we demonstrate that a dc electric field can induce a non-zero Berry curvature dipole through its coupling with the quantum metric in higher-wave-symmetric magnets, thereby producing a second-order anomalous Hall response in these systems. We study $p$, $d$, $f$, $g$, and $i$-wave symmetric magnets, a subclass of which ($d,g,i$) are altermagnets that break TRS. In all these systems, symmetry forbids the first- and second-order anomalous Hall responses by enforcing the vanishing of the first and second moments of the inherent Berry curvature of the occupied states. However, an external dc electric field that couples to the non-zero components of the quantum metric or Berry curvature polarizability can produce a correction to the Berry curvature distribution, which in turn generates a Berry curvature dipole. This BCD is highly tunable, scaling linearly with the dc field and having a non-trivial angular dependence. We predict the behavior of the induced BCD, and consequently, the second-order anomalous Hall conductivity, as a function of the angle $\phi$ between the dc electric field and the $x$-axis. We find that, among the unconventional magnets defined on the square lattice, the induced BCD and the nonlinear Hall response are qualitatively similar for $d$, $g$-wave (even-order) magnets, while exhibiting distinct behavior for $p$-wave (odd-order) magnets. Specifically, the angular dependence of the induced BCD and the second-order anomalous Hall response differs between these two groups. For instance, as shown in Fig.~\ref{fig:d_wave}(e), the induced BCD is always perpendicular to the dc electric field in the $d$ and $g$-wave magnets, independent of $\phi$, whereas, as illustrated in Fig.~\ref{fig:p_wave}(e), in the $p$-wave magnets, the induced BCD and the applied dc field are not always perpendicular to each other. 

Based on Eqs. (\ref{eq: secondOrderCurrent})$-$(\ref{eq:secondOrderAC}), Figs. \ref{fig:p_wave}(f), \ref{fig:d_wave}(f), and the analysis on page 5 of $p$-wave magnets and on page 6 of $d$-wave magnets, we conclude the following: for $d$-wave altermagnets, the driving ac electric field must have a component perpendicular to the dc electric field to generate a second-order anomalous Hall response transverse to the ac field through non-zero second-harmonic generation.
In contrast, for $p$-wave unconventional magnets, the second-order Hall response remains non-zero even when the ac and dc electric fields are parallel, as long as the dc field is applied at an angle $\phi \neq 0,\pi/2$ measured from the $x$-axis. Remarkably, the anisotropic nature of the field-induced BCD in the $p$-wave case allows for a second order ac anomalous Hall response that is highly tunable and a third order dc anomalous Hall response with a $\sin2\phi$ dependence (as seen in Fig.~\ref{fig:thirdOrder}). Thus, the electric-field-induced second- and third-order anomalous Hall effects in higher-order symmetric unconventional magnets can serve not only as a probe of the underlying quantum metric of the occupied states but also as a means to distinguish between even ($d$, $g$) and odd ($p$) order parameter symmetries defined on the square lattice. We found that the behavior of the $f$- and $i$-wave magnets on the triangular lattice is qualitatively similar to the behavior of $d$ and $g$-wave magnets on the square lattice.

The predicted signature of nonlinear Hall effect can be checked directly in experiments by using realistic systems that exhibit unconventional magnetic order. For example, candidate materials proposed to exhibit $p$-wave magnetic order are NiI$_2$ \cite{Song2025} and CeNiAsO \cite{hellenes2024pwavemagnets}. RuO$_2$ \cite{PhysRevB.99.184432,doi:10.1126/sciadv.aaz8809,Tschirner2023, Fedchenko2024,lin2024observationgiantspinsplitting} and Mn$_{5}$Si$_{3}$ \cite{PhysRevB.109.224430} are reported to host $d$-wave order. Possible candidates for $g$-wave magnetism include MnTe \cite{PhysRevLett.132.036702,Krempask2024} and CrSb \cite{Reimers2024,Yang_2025,Ding_2024,Cong_2024,Wan11}. For $f$-wave magnetism, Gd$_3$Ru$_4$Al$_12$ \cite{Hirschberger2019,Hirschberger2021} and Ba${_3}$MnNb$_2$O$_9$\cite{PhysRevB.90.224402} have been proposed, and $i$-wave magnetic order can be realized in twisted magnetic Van der Waals bilayers \cite{PhysRevLett.133.206702}.
\section{\label{sec: Acknowledgements}Acknowledgments} S.K. and S.T. acknowledge support from SC-Quantum, ARO Grant No. W911NF2210247 and ONR Grant No. N00014-23-1-2061. S.~N. also acknowledges financial support from Anusandhan National Research Foundation (ANRF), Government of India via the Prime Minister's Early Career Research Grant: ANRF/ECRG/2024/005947/PMS. 

\textit{Note added:} While preparing our manuscript, we noticed a preprint~\cite{Mukherjee_2025} that appeared recently on electric-field-induced Berry curvature dipole in $d$-wave altermagnets.

\bibliography{main} 
\end{document}